\documentclass[acmsmall,authorversion]{acmart}

\usepackage{listings}
\usepackage{glossaries}
\usepackage{booktabs, multirow}
\usepackage{algorithm} 
\usepackage{algorithmic}
\usepackage{balance}
\usepackage{rotating}
\usepackage{array}
\usepackage{enumitem}
\usepackage[normalem]{ulem}

\usepackage{tcolorbox}
\definecolor{mycolor}{rgb}{0.122, 0.435, 0.698}
\definecolor{gray1}{gray}{0.3}
\definecolor{darkgreen}{rgb}{0.0, 0.5, 0.0}
\definecolor{darkred}{rgb}{0.82, 0.1, 0.26}
\definecolor{shallowgreen}{RGB}{196, 214, 160}
\definecolor{shallowred}{RGB}{217, 149, 143}
\newcommand{\result}[1]{%
\begin{tcolorbox}[colframe=mycolor,boxrule=0.5pt,arc=4pt,
      left=6pt,right=6pt,top=6pt,bottom=6pt,boxsep=0pt,width=\columnwidth]%
      {#1}
\end{tcolorbox}%
}

\makeglossaries

\newglossaryentry{squality}
{
    name=SQuaLity,
    description={Our tool name}
}

\AtBeginDocument{%
  \providecommand\BibTeX{{%
    \normalfont B\kern-0.5em{\scshape i\kern-0.25em b}\kern-0.8em\TeX}}}

\setcopyright{rightsretained}
\acmJournal{PACMMOD}
\acmYear{2024} \acmVolume{2} \acmNumber{6 (SIGMOD)} \acmArticle{253} \acmMonth{12}\acmDOI{10.1145/3698829}

\makeatletter
\gdef\@copyrightpermission{
  \begin{minipage}{0.2\columnwidth}
   \href{https://creativecommons.org/licenses/by/4.0/}{\includegraphics[width=0.90\textwidth]{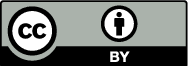}}
  \end{minipage}\hfill
  \begin{minipage}{0.8\columnwidth}
   \href{https://creativecommons.org/licenses/by/4.0/}{This work is licensed under a Creative Commons Attribution International 4.0 License.}
  \end{minipage}
  \vspace{5pt}
}
\makeatother
\begin{document}
\newcommand{\squality}{SQuaLity}
\newcommand{\artifacturl}{\url{https://doi.org/10.5281/zenodo.13896444}}
    
\newcommand{\NumTotalTestFiles}{4324}
\newcommand{\SQLiteTestFiles}{622}
\newcommand{\MySQLtestCommands}{112}
\newcommand{\SLTtestCommands}{4}

\newcommand{\SelfStatementKmeansClusters}{10}
\newcommand{\SelfQueryKmeansClusters}{10}
\newcommand{\SelfSampleNumber}{100}
\newcommand{\AcrossSampleNumber}{15}
\newcommand{\AcrossSampleNumberTotal}{100}
\newcommand{\SampleLimit}{30\%}

\newcommand{\NumFailingTests}{X}
\newcommand{\AvgSQLiteTestCases}{11907}
\newcommand{\AvgPostgreSQLTestCases}{147}
\newcommand{\AvgDuckDBTestCases}{15}
\newcommand{\AvgCockroachDBCases}{77}
\newcommand{\AvgRollbackProportion}{0.42\%}
\newcommand{\AvgBeginProportion}{0.62\%}
\newcommand{\AvgCommitProportion}{0.24\%}
\newcommand{\AvgWithProportion}{0.48\%}
\newcommand{\AvgExecuteProportion}{0.39\%}
\newcommand{\PostgreSQLGrantProportion}{0.65\%}
\newcommand{\DuckDBPragmaProportion}{6.99\%}
\newcommand{\PostgreSQLSetProportion}{3.62\%}
\newcommand{\SLTStandardTestFileProportion}{63.9\%}
\newcommand{\SLTCreateIndexFileProportion}{35.9\%}
\newcommand{\SLTStandardTestFileWithIndexProportion}{99.8\%}
\newcommand{\PostgresOverallStandardProportion}{68.9\%}
\newcommand{\LongtailStatementCount}{10.6\%}
\newcommand{\AvgStandardOverall}{25.7\%}
\newcommand{\AvgSimpleSelect}{79.9\%}
\newcommand{\AvgThreeSelect}{13.5\%}

\newcommand{\DuckDBSelfFailed}{5\%}
\newcommand{\PostgreSQLSelfFailed}{11\%}
\newcommand{\DuckDBEnvironmentIssues}{22}
\newcommand{\DuckDBPathsIssues}{22}
\newcommand{\PostgreSQLEnvironmentIssues}{88}
\newcommand{\PostgreSQLPathsIssues}{14}
\newcommand{\PostgreSQLConfigurationIssues}{7}
\newcommand{\PostgreSQLSetupIssues}{67}
\newcommand{\DuckDBExtensionIssues}{0}
\newcommand{\PostgreSQLExtensionIssues}{10}
\newcommand{\PostgreSQLClientIssues}{0}
\newcommand{\DuckDBClientIssues}{77}
\newcommand{\DuckDBFormatIssues}{58}
\newcommand{\SLTRunnerIssues}{2}
\newcommand{\DuckDBRunnerIssues}{1}
\newcommand{\PostgresqlRunnerIssues}{2}
\newcommand{\DuckDBNumericIssues}{17}
\newcommand{\DuckDBExceptionIssues}{2}
\newcommand{\SQLiteSkip}{19.80\%}
\newcommand{\DuckDBSkip}{39.47\%}
\newcommand{\PostgreSQLSkip}{3.12\%}
\newcommand{\DuckDBRequireFiltered}{26.2\%}
\newcommand{\EnvironmentSamplePercentage}{54\%}
\newcommand{\ExtensionSamplePercentage}{5\%}
\newcommand{\ClientSamplePercentage}{38\%}

\newcommand{\TotalErrorSLT}{}
\newcommand{\TotalErrorDuckDB}{}
\newcommand{\TotalErrorPostgreSQL}{}
\newcommand{\AvgUnSupSLT}{28}
\newcommand{\SLTFailureExecution}{16.6\%}
\newcommand{\DuckDBFailureExecution}{94.9\%}
\newcommand{\PostgreSQLFailureExecution}{96.0\%}
\newcommand{\SLTIncorrectResult}{X\%}
\newcommand{\DuckDBIncorrectResult}{X\%}
\newcommand{\PostgreSQLIncorrectResult}{X\%}

\newcommand{\PostgreSQLCompatibilityIssueSQL}{26}
\newcommand{\PostgreSQLCompatibilityIssueFunction}{7}
\newcommand{\PostgreSQLCompatibilityIssueType}{10}
\newcommand{\PostgreSQLCompatibilityIssueOperator}{6}
\newcommand{\PostgreSQLCompatibilityIssueConfiguration}{1}
\newcommand{\PostgreSQLCompatibilityIssueSemantic}{3}
\newcommand{\DuckDBCompatibilityIssueSQL}{30}
\newcommand{\DuckDBCompatibilityIssueFunction}{9}
\newcommand{\DuckDBCompatibilityIssueType}{5}
\newcommand{\DuckDBCompatibilityIssueOperator}{3}
\newcommand{\DuckDBCompatibilityIssueConfiguration}{1}
\newcommand{\DuckDBCompatibilityIssueSemantic}{7}
\newcommand{\DuckDBCompatibilityIssueMisc}{4}
\newcommand{\SQLiteCompatibilityIssueSQL}{9}
\newcommand{\SQLiteCompatibilityIssueType}{1}
\newcommand{\SQLiteCompatibilityIssueOperator}{2}
\newcommand{\SQLiteCompatibilityIssueSemantic}{5}

\author{Suyang Zhong}
\affiliation{%
  \institution{National University of Singapore}
  \country{Singapore}
  \orcid{0009-0003-0341-7362}
}
\email{suyang@u.nus.edu}

\author{Manuel Rigger}
\affiliation{%
  \institution{National University of Singapore}
  \country{Singapore}
  \orcid{0000-0001-8303-2099}
}
\email{rigger@nus.edu.sg}

\title{Understanding and Reusing Test Suites\\ Across Database Systems}
\renewcommand{\shorttitle}{Understanding and Reusing Test Suites Across Database Systems}



\begin{abstract}
Database Management System (DBMS) developers have implemented extensive test suites to test their DBMSs.
For example, the SQLite test suites contain over 92 million lines of code.
  Despite these extensive efforts, test suites are not systematically reused across DBMSs, leading to wasted effort.
  Integration is challenging, as test suites use various test case formats and rely on unstandardized test runner features.
 We present a unified test suite, SQuaLity, in which we integrated test cases from three widely-used DBMSs, SQLite, PostgreSQL, and DuckDB. In addition, we present an empirical study to determine the potential of reusing these systems' test suites.
 Our results indicate that reusing test suites is challenging: First, test formats and test runner commands vary widely; for example, SQLite has \SLTtestCommands{} test runner commands, while MySQL has \MySQLtestCommands{} commands with additional features, to, for example, execute file operations or interact with a shell.
 Second, while some test suites contain mostly standard-compliant statements (e.g., 99\% in SQLite), other test suites mostly test non-standardized functionality (e.g., 31\% of statements in the PostgreSQL test suite are non-standardized).
 Third, test reuse is complicated by various explicit and implicit dependencies, such as the need to set variables and configurations, certain test cases requiring extensions not present by default, and query results depending on specific clients.
  Despite the above findings, we have identified 3 crashes, 3 hangs, and multiple compatibility issues across four different DBMSs by executing test suites across DBMSs, indicating the benefits of reuse. 
  Overall, this work represents the first step towards test-case reuse in the context of DBMSs, and we hope that it will inspire follow-up work on this important topic.
\end{abstract}



\maketitle

\section{Introduction}

Database Management Systems (DBMSs) are large, complex software systems that are widely used by applications to store and retrieve data. DBMSs consist of various components, such as query processors~\cite{selinger1979access}, log managers~\cite{mohan1992aries}, storage engines~\cite{chang2008bigtable}, and optimization engines~\cite{wu2018towards,neumann2018adaptive}. Thus, DBMSs are usually large and complex. MySQL, one of the most popular open-source DBMSs, consists of about 3.5M lines of code (LOC). Even SQLite, a relatively lightweight relational DBMS, contains about 241K LOC. Consequently, DBMSs can be affected by bugs.

Unsurprisingly, DBMSs are typically well-tested. Researchers have developed various techniques to automate testing DBMSs and successfully found logic bugs~\cite{rigger2020finding,rigger2020detecting,rigger2020testing, slutz1998massive,liang2022detecting, tang2023detecting}, performance issues~\cite{jung2019apollo, liu2022automatic, ba2023finding}, crash bugs~\cite{sqlsmith, zhong2020squirrel}, and transaction bugs~\cite{kingsbury2020elle}.
Developers also test DBMSs using test suites.
These test suites typically consist of test cases as well as a test runner that can be used to execute the test cases and validate their results.
Most importantly, so-called \emph{end-to-end test suites} consist of SQL statements and test the DBMSs from a user perspective. 
For example, \texttt{sqllogictest}~\cite{sqllogic} is SQLite's test suite, which includes millions of SQL statements.
\textcolor{black}{A common reason for the prevalence of such SQL test suites is to avoid a lock-in regarding their own implementation details~\cite{youtube_duckdb_2020}.}

Despite developers' significant effort in developing these test suites, test cases are not systematically reused between DBMSs, wasting significant developer effort.
A recent interdisciplinary seminar on DBMS reliability stresses the importance of this issue and identified \emph{"a common testcase specification format and a test corpus that can be shared between DBMS engineering teams"} as one of the primary future challenges~\cite{bohm2022ensuring}.
We believe that a software engineering angle on this problem is promising, as test reuse has been a significant concern for the software engineering community.

Reusing test cases poses three challenges.
First, the test case formats designed by different DBMSs' developers are different, which prevents their direct reuse. Some DBMSs' test suites use SQL scripts to execute the engine and validate the results, while others rely on additional annotations or commands in the test files, for example, \texttt{for-loop}s to execute one statement multiple times.
Second, while SQL has been standardized by ANSI/ISO~\cite{ISO_IEC_9075}, in practice, DBMSs implement various SQL dialects that differ in syntax and semantics. Most DBMSs provide unique features. For test cases exercising such features, simply running them on other DBMSs would fail, because they would not implement these features. 
For example, PostgreSQL provides functions starting with ``pg\_'', which are typically system functions used for administrative tasks or system monitoring. Test cases containing these functions would likely result in errors when being executed on other DBMSs.
Third, results are usually inconsistent among different DBMSs, or even the same DBMS in different configurations, for the same SQL query. The reason could be floating-point precision differences, unstable query plans, or content discrepancies.

To tackle the above challenges, this paper describes a large-scale, systematic study that we performed on four DBMSs---SQLite, MySQL, PostgreSQL, and DuckDB---to investigate the reusability of DBMSs' test suites. To this end, we pose the questions below:

\begin{itemize}
    \item \emph{RQ1: What features do end-to-end DBMS testing frameworks provide?} As a first step, we determined whether the test formats and runners are sufficiently similar to enable reusing their tests.
    \item \emph{RQ2: What do test cases typically look like?} Next, we sought to understand the characteristics of different test suites' test cases with respect to, for example, what SQL statements they included. 
    A high percentage of SQL statements defined by the standard suggests potentially higher reusability.
    \item \emph{RQ3: What are the challenges of executing end-to-end tests?} We aimed to understand the effort required in developing test runners to execute the test suites. For example, we sought to understand whether executing the tests would be as simple as sending the SQL statements to the DBMS under test and validating its results, or whether any additional challenges would need to be accounted for.
    \item \emph{RQ4: Can test cases written for one DBMS find bugs in other DBMSs?} As our main goal was to investigate whether test suites can be reused, we sought to understand whether, assuming a common test format and runner, we would be able to successfully execute test cases written for one DBMS on other DBMSs. Furthermore, we aimed to investigate whether failing test cases could indicate overlooked bugs.
\end{itemize}

To answer the above questions, we studied DBMSs' test suites and derived a test format and test runners that allowed us to execute the test suites of SQLite, PostgreSQL, and DuckDB on any of these DBMSs; besides the insights of the empirical study, we present \squality{}, a unified test suite, which, to the best of our knowledge, represents the first step toward systematic test case reuse for DBMSs.

Our results demonstrate that test reuse can be useful, as we found new bugs, despite the challenges posed by the various test case formats and DBMS-specific features used in the test cases.
For RQ1, we found that test suites of DBMS used different formats of their test cases, and contained \SLTtestCommands{} to \MySQLtestCommands{} unique non-SQL commands interpreted by the test runners, for environmental settings or execution control. 
For RQ2, we found that \texttt{SELECT}, \texttt{INSERT} and \texttt{CREATE TABLE} are the most common SQL statements in the test suites, which indicates a high potential for reusability, as these statements are standardized.
However, we also found that test suites contained dialect-specific statements, posing difficulties for reuse.
For RQ3, we found that test cases required specific environments, extensions, and client dependencies to be tested appropriately.
For RQ4, we found 3 crashes as well as 3 hangs and reported the issues to the developers, overall 3 of which have been fixed. 
We also found suspicious discrepancies across SQL dialects, some of which we reported to the developers, which subsequently led them to discuss the intended semantics of the statements.

In summary, we make the following contributions:

\begin{itemize}
    \item an empirical study on the characteristic and reusability of different DBMSs' test suites (see Section~\ref{sec:Result-RQ1}--\ref{sec:Result-RQ3});
    \item \squality{}, a test suite that unifies the tests of SQLite, PostgreSQL, and DuckDB.
\end{itemize}

\section{Methodology}

In this paper, we sought to adapt the existing DBMSs' test suites to a common platform, to study both the opportunities and challenges of reusing the test suites, aiming to answer the above-posed research questions. We (1) extracted the test cases of each DBMS and converted them to a common format, (2) implemented a unified test runner to execute the test cases across DBMSs, and (3) analyzed the test cases we extracted in step (1) and the results of the test cases in step (2). We detail these steps in the following paragraphs. 


\paragraph{Terminology}\label{sec:term}
We specify the terms we use in the remainder of the paper. 
A \emph{test case} consists of an SQL statement and a specification of its expected behavior. A \emph{test file} contains several SQL \emph{test cases} and a \emph{test suite} consists of multiple test files.
Note that test cases in a test file might have implicit dependencies between each other. For example, one test case might check that a row was inserted successfully into a table, while another test case might validate a query's expected result, which depends on the successful execution of the \texttt{INSERT} statement.
The \emph{test runner} parses the \emph{test files}, executes the \emph{test cases}, and validates them.
In the subsequent paragraphs, the term \emph{test suite} is used to denote a combination of both the test runner and the collection of test cases, unless explicitly stated otherwise.
\emph{Failed test cases} refer to test cases that, when executed by the test runner, produce behavior that deviates from the expected outcome. 
Thus, the term refers to both statements that compute the expected result, or expectedly fail execution.

\begin{table}[tb]
\captionof{table}{DBMS rankings and their test suites information}\label{table:suties}
\setlength{\tabcolsep}{2pt}
\begin{tabular}{ lrrrrrrr }
\toprule
DBMS & DB-Engines & GitHub  & DBMS & Test Suite& Test\\
Names & Rankings & Stars  & Version & Version & Files\\
\midrule
SQLite & 9 & 4.5k & 3.41.1 & \texttt{a22803} & 622&\\ 
MySQL & 2 & 9.5k & 8.0.33 & \texttt{ea7087} & 1418 & \\
PostgreSQL & 4  & 13.2k  & 15.2 & \texttt{bc9993}& 212\\ 
DuckDB & 103 & 11.9k & 0.8.1 & \texttt{6536a7} & 2537 \\
\bottomrule
\vspace{-3mm}
\end{tabular}
\end{table}
    
\begin{figure}
  \centering
  \includegraphics[width=.7\linewidth]{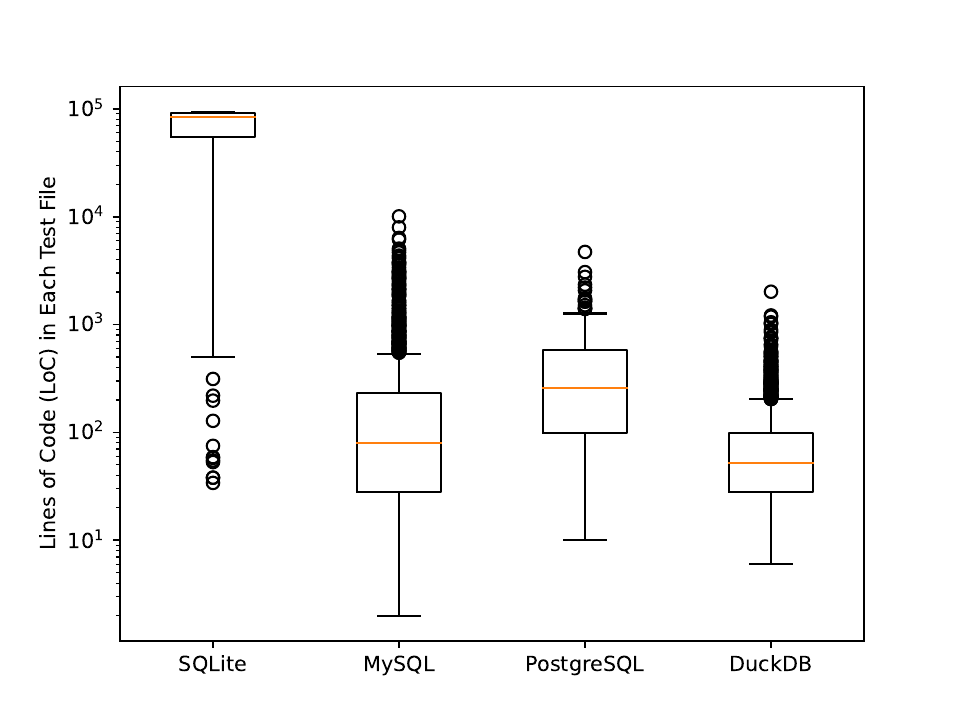}
  \caption{Number of test case lines per file of each DBMS (logarithmic scale).}\label{figure:test-case-length}
\end{figure}

\paragraph{Selecting DBMSs} 
We chose four popular open-source DBMSs, whose test suites we subsequently investigated (see Table~\ref{table:suties}).
We selected these DBMSs because they are highly popular based on ranking such as the DB-Engines ranking~\cite{DBEngines2023}, which ranks DBMSs according to their popularity, or their number of GitHub stars.
These DBMSs have also been the target of various automated testing works~\cite{rigger2020detecting,rigger2020finding,rigger2020testing,zhong2020squirrel, liang2022detecting, ba2023testing, sqlsmith, jiang2023dynsql, fu2023griffin}.
MySQL and PostgreSQL are traditional client-server DBMSs. 
SQLite and DuckDB~\cite{raasveldt2019duckdb} are embedded DBMSs, where the database system runs in the same process as the application it uses, making it particularly suitable for, for example, embedded devices. 
DuckDB is an Online Analytical Processing (OLAP) system, while the other three DBMSs are Online Transactional Processing (OLTP) systems. 

\paragraph{Selecting test suites}
From the four selected DBMSs, we identified four test suites, SQL Logic Test (SLT) for SQLite, MySQL Test Framework~\cite{MySQL2023}, the PostgreSQL regression test, and the DuckDB test suite, as shown in Table \ref{table:suties}.
Many DBMSs use several test suites for different purposes, implemented in different programming languages. 
All four DBMSs mentioned above have comprehensive SQL test suites, with detailed documentation and a large number of test cases (see Figure~\ref{figure:test-case-length}). 
In our study, we considered only test suites written in SQL, as we sought to reuse the test suites, and each DBMS uses SQL as a primary interface. 
Differences between DBMS implementations would make it difficult to directly reuse test cases consisting of API calls.
For example, test cases from DuckDB written using the C API, necessitate recompilation and relinking with every modification in the common header. This process can be time-consuming, particularly when the test suite is extensive~\cite{DuckDBSlides}.
Besides, the internal APIs that other DBMSs provide would significantly differ.
We omitted these non-SQL test suites, as they could not be easily reused. 
We disregarded commercial DBMSs like Oracle or SQL Server, as the test suites and runners are not publicly available. 


\paragraph{\squality{}}
\sloppy{}
We built a unified test suite, \squality{},\footnote{Available at \artifacturl{}.} to answer RQ1--RQ4. Its core logic is implemented in about 3,000 LOC to support analyzing test cases and executing them across different DBMSs. 
\squality{} can parse test files from each DBMS into individual SQL statements and extract the test runner commands.
The unified format is currently designed as an internal intermediate representation.
Thus, we do not consider the format as a core contribution of this work and for brevity, omitted describing it.
To execute test cases and obtain the test case results, \squality{} uses the Python DBMS connectors, allowing it to compare the DBMSs' results in a consistent manner.
If we had used the \emph{Command Line Interface} (CLI) clients, we would have had to parse and convert the CLI results returned in text format, which differs for each DBMS.
\squality{} executes and validates the test cases in a statement-by-statement manner. 
We omitted support for uncommon test suite features to limit the implementation effort to a reasonable degree---one of the authors was focusing on the project for a period of 12 months.

\paragraph{Methodology RQ1} 
We systematically analyzed and identified the common features of each test suite, with a focus on test case formats and test runner functionalities.
We conducted three analyses, in each of which we could refine the results and insights of the previous analyses.
First, we collected the test cases and documentation of each test suite. In this step, we analyzed how the test runner interprets elements of test cases, including how it parses the test files, executes the test runner commands, and validates the test results.
Second, we analyzed the commonalities and differences between these test suites, which are discussed in Section~\ref{sec:RQ1-testrunnercommands}. Based on the findings, we implemented \squality{}.
Third, we parsed and executed the test cases using \squality{}, which converts various test case formats into an internal unified format and identifies test runner commands. This, in turn, allowed us to refine and validate our findings from step one.


\paragraph{Analyzing the test cases}
We sought to investigate the test cases at the granular level of individual SQL statements, which helped us answer the subsequent RQs.
However, extracting individual SQL statements from the test cases is challenging, as the SQL statements are embedded in comments, test runner commands, and the specified expected results. 
A naive approach, for example, using regular expressions, would lead to false positives and negatives due to the complexity of SQL syntax.
Besides, distinguishing different SQL statement types is difficult, considering that we analyzed test cases from different DBMSs with different dialects.
As we sought to understand whether the test cases are suitable for execution across DBMSs, we adopted a best-effort approach.
We obtained the test cases from each DBMS test suite by implementing a corresponding parser that obtained each individual runner command and SQL statement within each test file.
Then, we relied on \texttt{sqlparse},\footnote{\url{https://github.com/andialbrecht/sqlparse}} a best-effort SQL-dialect agnostic parser, to identify the type of each individual SQL statement. 

\paragraph{Methodology RQ2}
We investigated the test contents by analyzing the individual SQL statements for SQLite, PostgreSQL, and DuckDB.
We omitted the MySQL test suite because its complex test case format and numerous runner commands make it highly specific to MySQL, limiting its potential reuse.
We wanted to determine how SQL is used in the three test suites, because this could answer whether the test cases are suitable for reusing. 
To this end, we first analyzed whether the test cases use SQL statements whose syntax is defined in the SQL standard~\cite{ISO_IEC_9075}---we subsequently refer to such statements as \emph{standard compliant}.
We categorized all the SQL statements according to their statement type (\emph{e.g.}, the statement type of \texttt{SELECT * FROM t0} is \texttt{SELECT}) and observed the distribution of different SQL commands usage as well as whether they are standard-compliant.
While  the analysis was on a statement level,  statements could still contain dialect-specific functions and keywords; for example, while in \texttt{SELECT to\_json ( date ' 2014 -05 -28 ');}, the \texttt{SELECT} statement is standardized, it references a PostgreSQL-specific JSON function.
We addressed this limitation in RQ4 by investigating whether the DBMSs could successfully produce the expected results for the statements.


\paragraph{Understanding the testing environment}
We observed that the unified test suites need to carefully consider the state of the DBMS and environment to successfully execute the test cases. 
A naive implementation of a test runner would pass the SQL statements to the DBMSs, but we found that various \emph{dependencies}---certain pre-execution set-up and requisites of the DBMS and the environment---need to be considered to produce the intended results.
Each DBMS has unique configurations, and implementing a new test runner could lead to missing configurations (or misconfigurations), producing unexpected results that are difficult to detect~\cite{do2013xu}.
The documentation of the PostgreSQL test suite mentions the above issue.\footnote{\url{https://www.postgresql.org/docs/15/regress-evaluation.html}}
Besides, considering that DBMSs allow interactions through various clients---CLIs and database connectors often provided for various programming languages---the risk exists that clients might present results in different ways.
Test cases that are strongly dependent on such dependencies could hardly be reused across DBMSs. 

\paragraph{Methodology RQ3} \label{sec:method-RQ3}
We sought to systematically study the dependencies of each test suite as they complicate the implementation of test runners.
Conceptually, we \emph{transplanted}~\cite{automated2015barr} the test suites into \squality{}'s test corpus, that is, by extracting and parsing test cases from the test suite of each DBMS into a format that \squality{} supports.
We executed and validated the transplanted test suites on the \emph{donor}---the DBMS for which the test suites were originally designed.
We collected the test results, including \emph{passed test cases}, for which the actual behavior matched the expected behavior specified by the corresponding test cases, and conversely, \emph{failed test cases}.
Then, we randomly sampled---following the methodology of other studies~\cite{baltes2022sampling}---\SelfSampleNumber{} failed test cases per DBMS to investigate the dependencies of each test suite.
We manually re-executed the test cases using the client of the donor test suite to determine if they were dependent on specific clients. Additionally, we examined the documentation of each donor to investigate potential environment and extension-related issues.

\paragraph{Methodology RQ4}
We sought to understand whether test cases written for one DBMS (\emph{donor}) could expose issues in another DBMS (\emph{host}). 
We executed test suites from different donors across DBMSs using our unified test runner and compared the actual results with the expected results. 
We analyzed the failed test cases, which could be caused by SQL dialect differences, settings discrepancies, or potential bugs. 
We exhaustively investigated the failures in SLT and applied the same sampling methods used in RQ3 to the other test suites.
For SLT, since it contains mostly standard-compliant SQL commands and error types are relatively simple, we could investigate all the failed test cases.
Conversely, other test suites exhibit various failures, and a thorough investigation would cost much effort.
To address this, for the results of each test pair (\emph{e.g.}, executing SQLite on the DuckDB test suite), we randomly sampled a test case and designed a rule that could identify other cases sharing similar patterns with it, subsequently filtering the remaining cases. 
We iteratively applied this procedure until either all cases could be classified according to the established rules, or the number of rules had reached an arbitrary, but reasonable, predefined limit of \AcrossSampleNumber{}. 
When the \AcrossSampleNumber{} rules could not classify all the unexpected results, we sampled the execution results. 

\paragraph{RQ4 Failure investigation}
As part of RQ4, we systematically analyzed the root causes for the failures, similar to a recent study on transaction incompatibilities~\cite{cui2023differentially}.
We analyzed a failed test case in the following steps.
First, we isolated the SQL statements from the test file, reduced them~\cite{zeller2002simplifying}, and re-executed them. 
Second, we manually investigated the documentation of each DBMS for an explanation of the observed discrepancy. We classified the discrepancy as a compatibility issue, where the behavior was documented by the developers.
Third, if the documentation failed to explain the discrepancies, we reported them to the developers.

\section{Test Suite Features (RQ1)}\label{sec:Result-RQ1}

\paragraph{Overview}\label{sec:RQ1-testcaseformat}
We investigated four test suites---SLT, the MySQL Test Framework, the PostgreSQL regression tests, and the DuckDB test suite---from two perspectives: test case formats and test runner commands.
Test cases differ in how they are structured in test files and how expected results are represented.
Additionally, different runners use various non-SQL commands, rather than only sending SQL statements to the DBMS and validating their results.


\paragraph{Test file formats}
The test file format varies based on whether the SQL statements and result specifications are explicitly separated in each file.
DuckDB specified their test cases in the SLT format, with some minor differences from the original SLT format as detailed below.
Each test case of SLT is an independent file consisting of several records, usually SQL statements with a header that specifies the expected behavior, as shown in Listing~\ref{listing:example-slt}. 
In this example, two statements create a table \texttt{t1} and insert three rows, which are expected to execute successfully. 
Then, a \texttt{SELECT} query fetches the data from the table and uses a \texttt{rowsort} comparator to validate the results, which is specified after the ``\lstinline{----}''. We discuss the query result format in the subsequent paragraph.
For each record, its SQL statement is executed individually to validate if the DBMS computes the correct result~\cite{sqllogic}. 
Conversely, the test cases of MySQL and PostgreSQL are not explicitly split in each test file, that is, the test runner executes the entire test file all at once. Then, it only compares whether the output of the entire test file matches the expected result.
Listing~\ref{listing:example-mysql} shows an example of a MySQL test. 
Each MySQL test case is a pair of two files, a test file, and a result file. A test file is a series of SQL statements and test runner commands. A result file is a copy of the test file, with the expected results after each SQL statement. 




    
\begin{lstlisting}[language=SQL, float=tb, caption={SQLite tests typically consist of a single file, where each SQL statement is annotated by its type and expected effect/result}, label={listing:example-slt}, belowskip=0pt]
statement ok
CREATE TABLE t1(a INTEGER, b INTEGER, c INTEGER)

statement ok
INSERT INTO t1(c,b,a) VALUES (3,4,2), (5,1,3), (1,6,4)

query I rowsort
SELECT a, b FROM t1 WHERE c > a;
----
2
4
3
1
\end{lstlisting}
\begin{lstlisting}[ language=SQL,float=tb,  caption={MySQL tests typically consist of two files: a test file that contains SQL and test runner commands, and a result file that contains the expected output from the runner executing the test file.}, label={listing:example-mysql}, belowskip=0pt]
# t/example.test
CREATE TABLE t1(a INTEGER, b INTEGER, c INTEGER);
INSERT INTO t1(c,b,a) VALUES (3,4,2), (5,1,3), (1,6,4);
SELECT a, b FROM t1 WHERE c > a;
# r/example.result
CREATE TABLE t1(a INTEGER, b INTEGER, c INTEGER);
INSERT INTO t1(c,b,a) VALUES (3,4,2), (5,1,3), (1,6,4);
SELECT a, b FROM t1 WHERE c > a;
a b
2 4
3 1
\end{lstlisting}

\paragraph{Result formats}
The query results specification for each test case varies across different test suites.
Results in SLT test cases are in value-wise ordering, in which each value of the query must appear on a separate line, as shown in Listing~\ref{listing:example-slt}.
DuckDB uses both value-wise ordering, as well as row-wise ordering, the latter of which resembles the structure of a table. Each line in the expected result represents a table row.
Listing \ref{listing:example-duckdb} shows an example in which the result is in row-wise order.
MySQL uses test results in a row-wise format, with an additional header to denote the column names.
PostgreSQL results, returned by the CLI, are strings representations of the table, containing the table header, values, and row count.
All formats convey information about table values.
Thus, lossless information transfer between the formats is possible.

\paragraph{Non-SQL commands overview}\label{sec:RQ1-testrunnercommands} 
We listed the Non-SQL commands in Table~\ref{table:RQ1-testrunner-feature} and detail them in the next paragraphs. ``$\checkmark$'' indicates that the corresponding DBMS's test suite supports the feature, while the number indicates the total number of supported commands.
Test suites can implement the same functionalities using different names; we grouped them together in the table.
The MySQL test suite contains the most runner commands, making it difficult to reuse its test cases.
Conversely, SLT supports only few runner commands. DuckDB's test suite format is based on SLT, but provides additional functionality.
PostgreSQL test cases are SQL scripts including psql commands. Psql\footnote{\url{https://www.postgresql.org/docs/current/app-psql.html}} is the CLI for PostgreSQL and is used to execute the PostgreSQL test cases.

\paragraph{Environmental settings}
Commonly, test runners support commands to set up the testing environment. 
\emph{Include} is used to read test code from other files, which enables developers to extract contents shared by multiple test files.
\emph{Set Variable} is used to set a specific variable that may be used in the following test procedure.
\emph{Load} loads data or custom functions from a specific source location. 

\begin{lstlisting}[language=SQL,float=tb,  caption={DuckDB test results are usually in row-wise format: each line is a row of the result table },label={listing:example-duckdb}, belowskip=0pt]
statement ok
CREATE TABLE t1(a INTEGER, b INTEGER, c INTEGER)

statement ok
INSERT INTO t1(c,b,a) VALUES (3,4,2), (5,1,3), (1,6,4)

query I
SELECT a, b FROM t1 WHERE c > a;
----
2 4
3 1
\end{lstlisting}
\begin{lstlisting}[language=SQL, float=tb,  caption={In MySQL, the "/" operator consistently performs floating-point division, even if both operands are integers. The "DIV" operator should be used if an integer division is needed.},label={listing:example-skiptest}, belowskip=0pt]
onlyif mysql # DIV for integer division: 
query I rowsort label-11
SELECT ALL 62 DIV ( + - 2 )
----
-31

skipif mysql # not compatible
query I rowsort label-11
SELECT ALL 62 / ( + - 2 )
----
-31
\end{lstlisting}
\begin{table}[tb]
    \small
\centering
\caption{Non-SQL commands of each DBMS test runner}\label{table:RQ1-testrunner-feature}
\vspace{-3mm}
\begin{tabular}{ lcccc}
\toprule
 Feature & SQLite & MySQL & PostgreSQL &  DuckDB \\
\midrule
Include & - & \checkmark & \checkmark &  - \\ 
Set Variable & \checkmark & \checkmark & \checkmark &  \checkmark \\
Load & - & \checkmark & \checkmark &  \checkmark \\
Loop & - & \checkmark & - &  \checkmark \\ 
Skiptest & \checkmark & - & \checkmark & \checkmark \\
Multi-Connections & - & \checkmark & \checkmark &  \checkmark \\
\midrule
CLI Commands & - & - & 114  & - \\ 
Runner Commands & 4 & 112 & - & 16 \\
\bottomrule
\vspace{-5mm}
\end{tabular}
\end{table}

\paragraph{Execution flow control}
Test runners support commands to manage the execution of SQL statements, allowing for more control than just sequential execution.
\emph{Loop} is used to execute the specific commands several times.
\emph{Skiptest} is used to conditionally skip the execution of tests. For example, in SLT, developers used these commands (see Listing~\ref{listing:example-skiptest}) to implement test cases where the SQL syntax varies from one DBMS to another. As another example, in DuckDB, \texttt{require sqlsmith} is used to check whether the current DuckDB has loaded the \texttt{sqlsmith} extension, which is a popular fuzz testing tool; if not, subsequent statements are skipped.
Some test runners support \emph{Multi-Connections}, enabling several connections to one database at a time.

\paragraph{Psql CLI commands}
As mentioned above, the PostgreSQL regression test suite uses the CLI to execute the tests, which provides meta-commands or shell-like features. In total, the suite makes use of 59 out of 114 unique CLI commands. 
Certain CLI commands have corresponding SQL commands in other DBMSs, for example, \texttt{\textbackslash c testdb} is equivalent to \texttt{USE testdb}, which changes the current database to \texttt{testdb}.
Some commands handle client-level functions, for example, \texttt{\textbackslash setenv} is used to set an environmental variable.
We did not seek to interpret and implement these commands; they are processed by the client of the DBMS, not the test runner.

\paragraph{MySQLTest commands}
The MySQL Test Framework supports \MySQLtestCommands{} unique runner commands.
Besides the features listed in Table~\ref{table:RQ1-testrunner-feature}, the test runner of MySQL supports various commands for local file operations (\emph{e.g.}, \texttt{writefile \textcolor{gray}{filename}}), server monitoring (\emph{e.g.}, \texttt{shutdown\_server [\textcolor{gray}{timeout}]}), and shell interaction (\emph{e.g.}, \texttt{exec \textcolor{gray}{command}}).
This enhances the test runner's ability to test the MySQL server, but the large amount of runner commands poses challenges to reusing the test cases.



\begin{figure*}[tb]
  \centering
  \includegraphics[width=\linewidth]{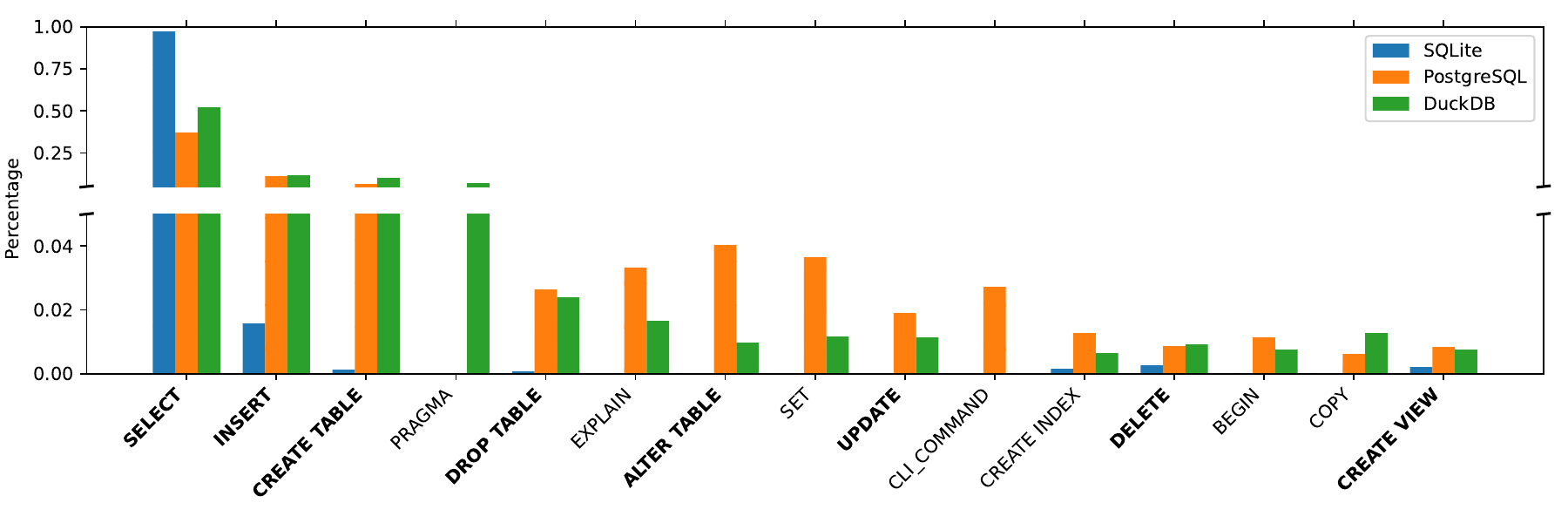}
  \caption{Distribution of SQL statement types in each DBMS test suite, with standard-compliant SQL statements denoted in bold.}\label{figure:RQ2-test-cases-patterns}
  \Description{SQL Pattern Distribution}
\end{figure*}

\result{
The SQLite, PostgreSQL, and DuckDB test suites use easily parsable test formats, whereas MySQL's cases embedded runner commands, which complicates reusability. 
Besides, the MySQL test runner's support for \MySQLtestCommands{} runner commands (\emph{e.g.}, compared to \SLTtestCommands{} in SQLite), and PostgreSQL's use of CLI commands both pose challenges to reuse.
}


\section{Test Case Patterns (RQ2)}\label{sec:SQL-usage}\label{sec:Result-RQ2}
We analyzed the SQL test cases, primarily by quantifying the frequency of each type of SQL statement present within the test suites. 
Figure~\ref{figure:RQ2-test-cases-patterns} shows the percentage of SQL statements in the three selected DBMSs' test suites. 
We ordered the SQL statements based on their mean occurrence frequency across three different DBMSs and selected the 15 highest-ranking ones for presentation.

\begin{table}[tb]
    \captionof{table}{The percentage of standard-compliant SQL statements among the test cases}\label{table:RQ2-test-cases-standard-compliant}
\begin{tabular}{lrr}
\toprule
DBMS & Overall & Exclusive  \\
Test Suite & Standard SQL & Standard Files\\
\midrule
SQLite & 99.76\% & 63.92\% \\
PostgreSQL & 68.89\% & 10.37\% \\
DuckDB & 76.14\% & 16.24\% \\
\bottomrule
\end{tabular}

\end{table}

\paragraph{Overview}
As shown in Table~\ref{table:RQ2-test-cases-standard-compliant}, SLT is the most standard-compliant test suite. Almost all test cases consist of standard-compliant SQL statements. 
However, only \SLTStandardTestFileProportion{} of the SLT test files contain only standard-compliant test cases, because \SLTCreateIndexFileProportion{} of the test files contain \texttt{CREATE INDEX} statements, used to create an index on columns to improve the speed of data retrieval~\cite{graefe2011modern}, which are not specified in the ANSI/ISO standard, despite being widely supported by DBMSs. 
If we counted \texttt{CREATE INDEX} as a standard-compliant statement, a higher standard compliance, that is, \SLTStandardTestFileWithIndexProportion{} of the test files contain standard-compliant test cases, could be achieved.
The proportion of standard SQL statements in PostgreSQL is \PostgresOverallStandardProportion{}, which is the lowest.


\paragraph{Common SQL statements}
As shown in Figure~\ref{figure:RQ2-test-cases-patterns}, all test suites contain common and standard-compliant SQL statements, such as \texttt{SELECT}, \texttt{CREATE}, and \texttt{INSERT}, demonstrating the potential of reusing the test cases.
\texttt{SELECT} allows querying data from the database. It is the most used SQL statement among all test suites.
\texttt{CREATE} is one of the Data Definition Language (DDL) statements in SQL, and it is primarily used to create tables. 
\texttt{DROP} and \texttt{ALTER} are also DDL statements that remove or modify the tables the previous \texttt{CREATE} statements create. 
Many other statements (\emph{e.g.}, \texttt{SELECT} and \texttt{INSERT}) operate on the table or a database object that was created using the \texttt{CREATE} statement.
\texttt{INSERT} is used to insert data into a table. It is frequently used following a \texttt{CREATE TABLE} statement. 
\texttt{UPDATE} and \texttt{DELETE} are used to modify and remove the data in the table, respectively.
While SQLite has the largest test suite SLT, with \SQLiteTestFiles{} test files, and each test file has \AvgSQLiteTestCases{} SQL statements on average, it contains only such fundamental SQL statements as above. 
The prevalence of standard SQL statements in SLT indicates a higher likelihood of being suitable for reuse.

\begin{lstlisting}[language=SQL, float=tb, caption={By altering the \texttt{explain\_output} setting to \texttt{OPTIMIZED\_ONLY}, the display changes from the default physical plan to present the optimized logical plan.},label={listing:example-pragma}, belowskip=0pt]
CREATE TABLE integers(i integer, j integer, k integer);
INSERT INTO integers VALUES (5, 5, 5), (10, 10, 10);
PRAGMA explain_output = OPTIMIZED_ONLY;
EXPLAIN SELECT k FROM integers where j=5;
\end{lstlisting}

\paragraph{Unstandardized SQL statements}
The test suites of PostgreSQL and DuckDB, despite having a smaller number of test cases, contain a broader range of statements, including SQL statements related to transactions, query plans, and DBMS-specific features.

Statements such as \texttt{SET} in PostgreSQL (\PostgreSQLSetProportion{}), and \texttt{PRAGMA} in DuckDB (\DuckDBPragmaProportion{}) configure the DBMS. Neither of the statements is specified by the SQL standard.
\texttt{SET} is used for changing session-level settings. 
\texttt{PRAGMA} is specific to SQLite and DuckDB.
For example, DuckDB uses a \texttt{PRAGMA} statement to change how query plans are displayed, as shown in Listing~\ref{listing:example-pragma}.
These statements inhibit the reuse because they are not standard-compliant and may not execute successfully across different DBMSs.
DBMSs like SQLite silently ignore unknown parameters specified by \texttt{PRAGMA}.
Subsequently, unsupported settings can lead to unexpected results in those test cases relying on them.


In addition to \texttt{SELECT} statements, test suites also apply other non-standard queries to validate specific properties of the system. 
\texttt{EXPLAIN} statements are used in PostgreSQL and DuckDB test suites to check the query plan correctness. 
DuckDB uses these statements to validate whether optimizations are expectedly performed~\cite{DuckDBSlides}.
Listing~\ref{listing:example-pragma} shows SQL statements from DuckDB test cases. The \texttt{EXPLAIN} statement retrieves the optimized logical query plan, which describes how the DBMS retrieves data from the table \texttt{integers} and filters based on the specified predicate.
Test cases using the \texttt{EXPLAIN} statement are unlikely to be reusable, because \texttt{EXPLAIN} is not a standard SQL statement, and the result formats of query plans differ between DBMSs~\cite{ba2024exploratory}.

\texttt{BEGIN} and \texttt{ROLLBACK} are statements that relate to transactions. Although \texttt{START TRANSACTION} is the standard SQL statement to start a transaction, we observed that these test suites more frequently use \texttt{BEGIN}. 
SQLite even lacks support for the standard syntax.
Typically, a transaction is either committed using \texttt{COMMIT} (\AvgCommitProportion{}) or aborted using \texttt{ROLLBACK} (\AvgRollbackProportion{}).
Considering that \texttt{ROLLBACK} and \texttt{COMMIT} are standard SQL, less effort is needed to reuse these test cases. 

\paragraph{Infrequently used SQL statements}
An average of \LongtailStatementCount{} statements are infrequently used in the test suite and thus not shown in Figure~\ref{figure:RQ2-test-cases-patterns}, such as \texttt{WITH} statements (\AvgWithProportion{}) to define a \emph{Common Table Expression} (CTE) and \texttt{EXECUTE} statements (\AvgExecuteProportion{}) to run a stored procedure or prepared statement.
We also found non-existent types of SQL statements.
First, developers specified intentionally incorrect statements to test the parser of DBMS, and our analyzer did not attribute these statements to the correct type (\emph{e.g.}, \texttt{SELEC} in DuckDB test cases).
Second, as we used a best-effort parser to analyze the SQL statement, certain corner cases (\emph{e.g.}, \texttt{(((((select * from int8\_tbl)))))} would be classified as \texttt{(((((SELECT}). However, these instances are rare, accounting for less than 0.1\% of the total test cases, and can therefore be considered negligible.

\begin{figure}[tb]
  \centering
  \includegraphics[width=.7\linewidth]{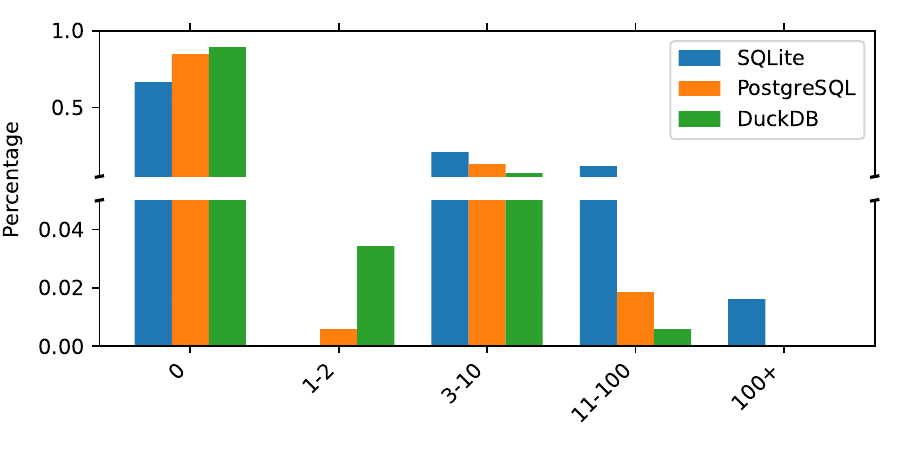}
  \caption{Distribution of tokens in \texttt{WHERE} predicates of \texttt{SELECT} statements. 0 shows \texttt{SELECT} statements without \texttt{WHERE} clause.}\label{figure:RQ2-select-patterns}
  \Description{SELECT Pattern Distribution}
\end{figure}

\begin{lstlisting}[language=SQL, float=tb, caption={Examples of \texttt{SELECT} statements in test suites},label={listing:example-select},belowskip=0pt]
SELECT interval '1-2'; -- 0 token
SELECT a, b FROM t1 WHERE c > a; -- 3 tokens
SELECT unit.total_profit FROM unit, unit2; -- Implicit join
SELECT a, test.b, c FROM test INNER JOIN test2 ON test.b = 2 ORDER BY c; -- Inner join
\end{lstlisting}

\textcolor{black}{
\paragraph{\texttt{SELECT} query complexity} 
In terms of \texttt{SELECT} statements, we found that most are rather simple. 
This is demonstrated in Figure~\ref{figure:RQ2-select-patterns}, which shows the number of tokens in \texttt{WHERE} predicates. 
Most queries, namely \AvgSimpleSelect{}, lack \texttt{WHERE} clauses, such as the example query shown in line 1 of Listing~\ref{listing:example-select}.
Based on our understanding of inspecting the test cases, queries without \texttt{WHERE} clauses were often used to test the specific functions and operators by evaluating them on constants, without referencing any tables, or to fetch and validate the data from the table after specific operations. 
On average, \AvgThreeSelect{} of queries contain 3 to 10 tokens in the predicates, such as the query shown in line 2 of Listing~\ref{listing:example-select}.
1.6\% of SLT query predicates consist of more than 100 tokens, indicating the use of complex expressions.
We also analyze the complexity of \texttt{JOIN} usage. 
Only 7.2\% of queries consist of either implicit (\emph{e.g.}, line 3 in Listing~\ref{listing:example-select}) or explicit joins (\emph{e.g.}, \texttt{INNER JOIN}, \texttt{LEFT JOIN} and \texttt{RIGHT JOIN}). 
Specifically, 5.1\% of queries contain implicit joins and 1.1\% of queries contain \texttt{INNER JOIN} in the three test suites.
}
\result{
The results suggest that DBMSs typically test DBMS-specific functionality, such as illustrated by more than 80\% of test files in PostgreSQL and DuckDB containing non-standard SQL statements. SQLite seems to be an exception, as 99.76\% of statements are standard-compliant.
}





\begin{table}[tb]
\captionof{table}{Running donor test suites against donor}\label{table:summary_self_results}
\begin{tabular}{ lrrrr }
\toprule
DBMS & Total & Executed & Failed  \\ 
Names & Test Cases & Cases & Cases \\ 
\midrule
SQLite & 7,406,130 & 5,939,879 & 2 \\ 
PostgreSQL & 36,677 & 35,534 & 4,075\\ 
DuckDB & 33,113 & 20,619 & 1,035 \\ 
\bottomrule
\end{tabular}
\end{table}


\section{Test Case Dependencies (RQ3)}\label{section:unified_test}\label{sec:Result-RQ3}




\paragraph{Overview}
Table~\ref{table:summary_self_results} shows the results of validation on the donor (\emph{e.g.}, running SLT on SQLite) using \squality{}.
\squality{} executed close to 6 million test cases from more than 7.4 million test cases that we collected.
The remaining tests included directives that caused them to be skipped.
Failed cases refer to test cases where SQL statements behave unexpectedly, as introduced in Section~\ref{sec:term}.
These were due to dependency issues, which indicate challenges for reuse. The detailed reasons are shown in Table~\ref{tab:table-self}, and we will discuss them in the subsequent paragraphs.



\begin{table}[tb]
    \caption{Classification of a sample of 100 failing test cases for each DBMS and its test suite}
    \label{tab:table-self}
    \small
\begin{tabular}{llrrr}
\toprule
 & Reason  & SQLite & DuckDB & PostgreSQL \\
\midrule
\multirow[c]{3}{*}{Environment} & File Paths & 0 & 22 & 14 \\
 & Setting & 0 & 0 & 7 \\
 & Set Up & 0 & 0 & 67 \\
\midrule
Extension & Extension & 0 & 0 & 10 \\
\midrule
\multirow[c]{3}{*}{Client} & Format & 0 & 58 & 0\\
 & Numeric & 0 & 17 & 0 \\
 & Exception & 0 & 2 & 0 \\
\midrule
Misc & Runner & 2 & 1 & 2 \\
\bottomrule
\end{tabular}
\end{table}

\paragraph{Pre-filtered test cases}
As shown in Table~\ref{table:summary_self_results}, \squality{} skipped \SQLiteSkip{}, \PostgreSQLSkip{}, and \DuckDBSkip{} of the test cases of SQLite, PostgreSQL, and the DuckDB test suite, respectively.  
SLT supports the \emph{skiptest} runner command, which can be used when the test case is DBMS-specific.
When testing against SQLite itself, some test cases were filtered intentionally, because they were designed to be executed on other DBMSs, and thus the test runner did not execute all the test cases.
Similarly, in DuckDB, the runner command \texttt{require} halts all the following test cases if one extension has not been loaded. 
PostgreSQL test cases are not explicitly separated, as mentioned in Section~\ref{sec:Result-RQ1}. Thus, we omitted test cases that were difficult to parse.

\paragraph{Environmental settings}
As shown in Table~\ref{tab:table-self}, in \DuckDBEnvironmentIssues{} out of the 100 sampled failed test cases in DuckDB, and in \PostgreSQLEnvironmentIssues{} for PostgreSQL, the test cases are dependent on specific environment settings.
The difference between the environment of developers and ours leads to discrepancies between the actual and expected results.
First, \DuckDBPathsIssues{} of the DuckDB test cases and \PostgreSQLPathsIssues{} of the PostgreSQL ones failed due to incorrect file path names that resulted in failures to load data, which affected the subsequent test cases. The data source was typically specified by hardcoded paths or paths set by environmental variables, and incorrect path names could result in no loaded data.
Second, \PostgreSQLConfigurationIssues{} PostgreSQL cases failed due to incorrect configurations.
DBMSs support locale settings to manage internationalization, which led to the difference between actual results and expected results.
Several other configurations including different system configurations, testing schema names, and default output format could also lead to discrepancies.
Third, \PostgreSQLSetupIssues{} failed PostgreSQL cases were due to the unsuccessful setup for the test database.
For example, PostgreSQL used a scheduler to execute specific test cases first to set up certain environment settings.
Conversely, test files in SQLite and DuckDB have no dependencies on each other and can be executed independently.
Test cases associated with environmental settings pose challenges for reusability, as different DBMSs use diverse default configurations and environment setting methods. 

\begin{lstlisting}[language=SQL, float=tb, caption={This \texttt{CREATE FUNCTION} statement is used to create a function in current database by loading C library from ``\texttt{regresslib}''.}, label={listing:example-postgresql-regress},belowskip=0pt]
CREATE FUNCTION test_opclass_options_func(internal)
    RETURNS void
    AS :'regresslib', 'test_opclass_options_func'
    LANGUAGE C;
\end{lstlisting}

\paragraph{Database engine extensions}
\PostgreSQLExtensionIssues{} of the failing PostgreSQL test cases require extensions to be loaded, such as the shared library \texttt{regress.so}, which exposes functions such as shown in Listing~\ref{listing:example-postgresql-regress}, which are used in the subsequent test cases.
Lack of or failure to load these functions could result in the failure of the current statement and subsequent test cases.
Besides, as previously noted, DuckDB test cases use the runner command \texttt{require} to explicitly state their required extensions, and the remaining test cases are not executed if the required extension is not loaded.
Thus, we encountered no failed test cases in our DuckDB samples due to this issue, since \DuckDBRequireFiltered{} of the cases have been pre-filtered.
Extension-related test cases inhibit reusing, as, typically, extensions are DBMS-specific.

\paragraph{Clients}
\DuckDBClientIssues{} of the failing DuckDB test cases rely on specific clients.
Our selected DBMSs support various client APIs, and the respective test runners are implemented using different client APIs, C API for SQLite, C++ for DuckDB, and psql for PostgreSQL. Certain result objects are differently presented depending on the client, which led to discrepancies. 
Listing~\ref{listing:example-client-diff} shows an example of result discrepancies under different clients.
The discrepancies mainly arise from variations in output formats, such as the representation of nested data types like lists and structs, binary objects, and numeric data, which affected \DuckDBFormatIssues{} DuckDB test cases.
Numeric data issues arise from the original test runner's lenient comparison, as our test runner demands an exact match with the expected results, given that it could provide consistency and catch subtle issues.
This affected \DuckDBNumericIssues{} DuckDB test cases.
As exemplified in Listing~\ref{listing:example-duckdb-numeric}, 
 the expected result listed in the test case is 4999, while the true expected answer should be 4999.5.
The discrepancies in result formats caused by different clients add to the difficulty of reusing the test cases.

Finally, we observed issues in clients, as shown in Listing~\ref{listing:client-exception}. A \texttt{Not Implemented Error} arose in the DuckDB Python interface, while the same statements were executed as expected in the CLI, despite both being executed on version \texttt{0.8.1}. 
We found one case whose result differs between the CLI and Python clients (see Listing~\ref{listing:example-order-diff}).
We have reported this issue to the developers,\footnote{\url{https://github.com/duckdb/duckdb/issues/5413}} after which they implemented tests for various client connectors.
These cases demonstrate that specific clients are required for some test cases.



\begin{lstlisting}[language=SQL, float=tb,  caption={Different results (after converting to string)}, label={listing:example-client-diff}, belowskip=0pt]
SELECT ARRAY[1,2,3,'4'];
-- Result in DuckDB
[1, 2, 3, 4]
-- Result in PostgreSQL
{1,2,3,4}
-- Result in DuckDB Python API
['1', '2', '3', '4']
\end{lstlisting}
\begin{lstlisting}[language=SQL, float=tb,  caption={Different results (after flattening values in a line) when executing the same query using the CLI and Python interfaces of DuckDB}, label={listing:example-order-diff}, ]
SELECT 1 UNION ALL SELECT * FROM range(2, 100) UNION ALL SELECT 999 LIMIT 5;
-- CLI: 1 2 3 4 5
-- Python interface: 1 999 2 3 4
\end{lstlisting}

\begin{lstlisting}[language=SQL, float=tb, caption={In DuckDB, floating-point results are considered matching if the difference is less than 1\%}, label={listing:example-duckdb-numeric},belowskip=0pt]
statement ok
create table quantile as select range r, random() from range(10000) union all values (NULL, 0.1), (NULL, 0.5), (NULL, 0.9) order by 2;

query I
SELECT median(r) FROM quantile -- Actual: 4999.5
----
4999
\end{lstlisting}
\begin{lstlisting}[language=SQL,float=tb,caption={Executing these statements using the DuckDB Python interface results in an error}, label={listing:client-exception},belowskip=0pt]
CREATE TABLE tbl1 (union_struct UNION(str VARCHAR, obj STRUCT(k VARCHAR, v INT)));
INSERT INTO tbl1 VALUES  ({'k': 'key1', 'v': 1});
SELECT * FROM tbl1; -- Expected: {'k': key1, 'v': 1}
\end{lstlisting}
\paragraph{Implications}
Omitted test cases from the donor systems were due to dependency issues related to environment, extension, and client, as shown in Table~\ref{tab:table-self}. 
Dependencies on these components are often related to DBMS-specific features and configurations that are not defined by the SQL standard and vary between DBMSs (\emph{e.g.}, one cannot use MySQL CLI on PostgreSQL).
As a result, reusing these test cases across different DBMSs is challenging and provides limited benefits.
We believe that, when designing a unified test suite for reuse, minimizing dependencies is crucial.
Each test case should be independent of others and free from DBMS-specific features.

\result{We identified three major kinds of test case dependencies concerning the environment, extensions, and clients. 
The SQLite test cases require few dependencies. 88\% of the failed PostgreSQL test cases were environment-related, and 77\% of the DuckDB ones relied on specific clients.
}

\section{Test Suite Compatibility (RQ4)} \label{sec:eva-rq4}
We executed the test cases from SLT, the PostgreSQL regression test suite, and the DuckDB test suite on SQLite, PostgreSQL, DuckDB, and MySQL.
First, we present the overall execution results of these test suites. Second, we show the crashes and hangs found during execution, which are excluded from the above results. Third, we investigate the failed test cases found during execution. 
All the test cases have been reduced~\cite{zeller1999yesterday} for presentation.

\begin{figure}[tb]
  \centering
  \noindent
  \includegraphics[width=.8\linewidth]{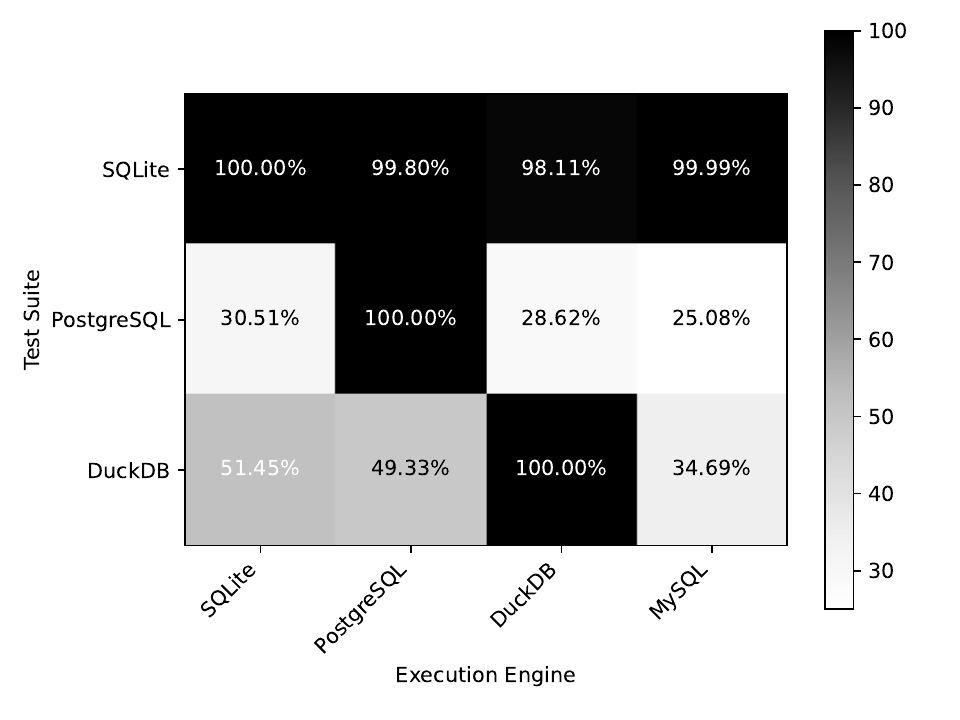}
  \caption{The percentage of SQL test cases that execute successfully. At each intersection of \emph{execution engine} and \emph{test suite}, the color represents the success rate with respect to the total number of test cases. Darker shades indicate a higher success rate. For example, 51.45\% of the DuckDB test cases are executed as expected on SQLite.}\label{figure:RQ4-test-results-success-rate}
\end{figure}
\begin{lstlisting}[language=SQL,float=tb,  caption={A crash in DuckDB}, label={listing:duckdb-crash-1}, belowskip=0pt]
ALTER SCHEMA a RENAME TO b;
\end{lstlisting}
\begin{lstlisting}[language=SQL,float=tb,  caption={A crash related to transaction handling in DuckDB}, label={listing:duckdb-crash-2}, belowskip=0pt]
CREATE TABLE a (b int);
BEGIN;
INSERT INTO a VALUES (1);
UPDATE a SET b = b + 10;
COMMIT;
UPDATE a SET b = b + 10;
\end{lstlisting}
\begin{lstlisting}[language=SQL, float=tb,  caption={A crash in a recursive CTE in MySQL}, label={listing:mysql-crash}, belowskip=0pt]
WITH RECURSIVE t(x) AS (
  SELECT 1
    UNION ALL
  (SELECT x+1 FROM t WHERE x < 4
    UNION
  SELECT x*2 FROM t WHERE x >= 4 AND x < 8)
) SELECT * FROM t ORDER BY x;
\end{lstlisting}
      

\paragraph{Overall execution}
Figure~\ref{figure:RQ4-test-results-success-rate} shows a heatmap that shows the execution results of running the test suites across different DBMSs, excluding test cases that crash or hang the DBMS engine.
SLT was the most compatible test suite, as all three other DBMSs passed over 98\% of its test cases. 
The PostgreSQL regression suite was the most incompatible one. It obtained an average success rate of 28.1\%.
The DuckDB test suite obtained an average success rate of 45.2\%.
These results were expected based on our analysis in Section \ref{sec:SQL-usage}, as the PostgreSQL regression suite contains the lowest percentage of standard-compliant SQL test cases and makes use of CLI commands, while SLT contains the most standard-compliant cases.
MySQL achieved a higher success rate on SLT compared to the other two test suites, because of the \texttt{skiptest} runner commands in SLT (see Section~\ref{sec:Result-RQ1}), which restricted the runner to execute only general test cases and those implemented in the MySQL SQL dialect.
We investigated the reason for the crashes and hangs, and examined the compatibility issues, which we explain in the next paragraphs.


\begin{lstlisting}[language=SQL, float=tb, caption={A recursive CTE led to an infinite loop}, label={listing:example-duckdb-hang}, belowskip=0pt]
WITH RECURSIVE x(n) AS (
  SELECT 1 UNION ALL 
  SELECT n+1 FROM x WHERE n IN (SELECT * FROM x)
) SELECT * FROM x;
\end{lstlisting}
\begin{lstlisting}[language=SQL, float=tb, caption={An eponymous table-value function in PostgreSQL triggered an overflow in SQLite}, label={listing:example-sqlite-hang},belowskip=0pt]
SELECT count(*) FROM 
generate_series(9223372036854775807,9223372036854775807);
\end{lstlisting}


\paragraph{Crashes} 
We found three crashes---unexpected terminations of the DBMS.
Listing~\ref{listing:mysql-crash} shows a recursive CTE that crashed the MySQL server when executing \texttt{FollowTailIterator::Read()}, which has been addressed. Given the critical nature of the bug, it was assigned a CVE (2024-20962). We discovered this crash using the DuckDB test suite.
Listing~\ref{listing:duckdb-crash-1} shows a test case leading to a crash in DuckDB \texttt{0.7.0}.
In the previous version (\texttt{0.6.1}), DuckDB correctly threw an \texttt{Not implemented Error}. 
Listing~\ref{listing:duckdb-crash-2} shows a crash that was caused by \texttt{UPDATE} after the \texttt{COMMIT} of the transaction.
We discovered these two crashes by executing the PostgreSQL regression test suite on DuckDB's latest release.
While DuckDB had already accumulated a large test suite, the fact that we encountered these crashes highlights the importance of test case reuse.

\paragraph{Hangs}
We observed three hangs---the DBMS entered an infinite loop when executing a statement or exhibited an overly-long execution time.
We found one of them when executing a query from SLT on MySQL, which joined more than 40 tables. This caused MySQL to take more than one minute to compile and determine an efficient join order in the default setting, \texttt{optimizer\_search\_depth = 62}. By replacing this value with 0, the query could be executed within one millisecond. 
Although this behavior was expected according to the MySQL documentation,\footnote{\url{https://dev.mysql.com/doc/refman/8.0/en/controlling-query-plan-evaluation.html}} the other three DBMSs could return results in a reasonable time with default settings. 
This suggests that users might need to change settings when migrating databases from other systems.
We found another hang when executing one PostgreSQL test (\texttt{with.sql}) on DuckDB, as shown in Listing~\ref{listing:example-duckdb-hang}. PostgreSQL and MySQL prevented this query from being executed by issuing an error, while it caused an infinite loop in DuckDB.
This hang reflects differences in the design philosophies behind the DBMSs.
Developers mentioned it is a deliberate design for DuckDB, and adding constraints is strongly against DuckDB's friendly SQL~\cite{Monahan_2022} because it would restrict users.
In addition, in a short, ad-hoc fuzzing campaign, in which we used the test suites as seed inputs, we discovered one hang (see Listing~\ref{listing:example-sqlite-hang}), which was confirmed as a bug\footnote{\url{https://sqlite.org/forum/forumpost/754e2d}} and subsequently fixed.
The hang was triggered when executing the
\texttt{generate\_series()} function in SQLite.
This function is a table-value function from PostgreSQL that SQLite implemented as an extension. 
This bug was introduced more than 3 years ago, which stresses the significance of test suite reuse, as it had been overlooked by other automated testing tools~\cite{zhong2020squirrel,rigger2020testing}.


\paragraph{Failed cases}
Failed test cases are due to (1) unexpected execution statuses and (2) unexpected query results.
When executing SLT on the three other hosts, \SLTFailureExecution{} of the failed test cases were due to unexpected execution statuses, and the percentage is \DuckDBFailureExecution{} and \PostgreSQLFailureExecution{} for the DuckDB and PostgreSQL test suites.
The reason for the higher execution success rate of SLT is that the SLT test cases contain mostly standard-compliant SQL statements, while the other two contain non-standard statements as well as DBMS-specific functions and data types.
For example, the \texttt{pg\_typeof()} function to obtain the data type of its argument is implemented by PostgreSQL and DuckDB, while no similar built-in function is supported in MySQL.
Conversely, failed test cases in SLT were mostly due to unexpected query results, while the other two contain only a few.
Certain functions or operators share the same name between DBMSs while the semantics are not the same, which mostly contributed to these mismatched query results. 
For example, when executing query \texttt{SELECT COALESCE(1, 1.0);}, SQLite returns the integer value \texttt{1}, PostgreSQL returns the floating-point value \texttt{1}, and both MySQL and DuckDB return the floating-point value \texttt{1.0}. However, all four DBMSs return the integer value \texttt{1} for \texttt{SELECT COALESCE(1, 1);}.

\begin{table*}[tb]
    
    \caption{Executing SQLite, DuckDB, PostgreSQL, and MySQL on the SLT, DuckDB test suite, and PostgreSQL regression test suite. We comprehensively analyzed the SLT test results, and randomly sampled and analyzed \AcrossSampleNumberTotal{} test cases from DuckDB test suite and PostgreSQL regression test suite. }
    \setlength{\tabcolsep}{4pt}
\footnotesize
\begin{tabular}{lrrrrrrrrr}
   \toprule
    Failed Reasons        & \multicolumn{3}{c}{SQL Logic Test} & \multicolumn{3}{c}{DuckDB Test Suite} & \multicolumn{3}{c}{PostgreSQL Test Suite}  \\
   Granularity  &  DuckDB &  PostgreSQL &  MySQL &SQLite &  PostgreSQL &  MySQL  &  SQLite &  DuckDB &  MySQL  \\
   \midrule
   Statements         & 1,317 & 4,905 & 915  &      37\%  &     25\%       &        41\%           &  54\%   & 53\%   & 58\%  \\
   Functions               & 0  & 0 & 0 &               34\%  &      26\%      &         18\%         &  11\%   & 10\%   & 8\%  \\
   Types                   & 43  & 12 & 0 &             13\%  &      36\%      &         22\%         &  15\%   & 23\%   & 16\%  \\
   Operators                & 7,075 & 6,069 & 0  &      14\% &     0\%    &        10\%               &  20\%      & 3\%    & 16\%   \\
   Configurations             &  0 & 0 & 0   &          0\%  &      3\%        &        4\%            &  0\%    & 8\%    & 2\%   \\
   Semantic                 & 104,033   & 609 & 0& 2\%  &       9\%     &        2\%             &  0\%    & 3\%    &  0\%   \\
   Misc                      &0 &0 &0                &   0\%  &        1\%       &         3\%         &  0\%    & 0\%    & 0\%   \\ 
   \midrule
   Timeout                            & 0      & 0      & 1 &   0      & 0      & 0 & 1      & 1      & 0    \\
   Crash                              & 0      & 0      & 0 &  0      & 0      & 1  & 0      & 2      & 0   \\ 
   \bottomrule
   \end{tabular}
    \label{tab:RQ4-Overall}
\end{table*}

\paragraph{Incompatibility issues}
We analyzed the root causes of the failed cases and categorized them into several types: unsupported (1) SQL statements, (2) functions, (3) types, (4) operators, as well as issues caused by (5) system settings, and (6) inconsistent semantics. We group together other reasons as a category miscellaneous, abbreviated as \emph{misc}. Note that one failed test case might be due to multiple issues. We selected the first error type that we identified---often indicated by an error message from the DBMS.
Table~\ref{tab:RQ4-Overall} shows the reasons for failed test cases. Each cell in the table represents the count or percentage of 100 samples of the failed cases. For example, when executing the DuckDB test suite on SQLite, 37\% of the samples were due to unsupported SQL statements.

The unsupported \emph{Statements} category refers to SQL statements from donor test cases that fail to execute on the host. These kinds of failures can occur for various reasons: either the host system lacks support for the statement, or there are differences in syntax and restrictions.
We found \SQLiteCompatibilityIssueSQL{}, \DuckDBCompatibilityIssueSQL{}, and \PostgreSQLCompatibilityIssueSQL{} of these kinds of issues when executing the SQLite, DuckDB, and PostgreSQL test suites, respectively.
Fewer test cases in SLT failed for this reason compared to those in the DuckDB and PostgreSQL suites, which aligns with the findings from RQ2.
First, non-standard SQL statements (\emph{e.g.}, \texttt{PRAGMA} and \texttt{SET}), which we mentioned in Section~\ref{sec:Result-RQ2}, led to most of the unexpected failures.
Second, DBMS-specific clauses in statements also led to failures, for example, one unique feature \texttt{ASOF\_JOIN} in DuckDB caused a failure.
Third, DBMSs may apply different constraints on specific operations; for example, different restrictions concerning the \texttt{WITH} statement, which we mentioned in the paragraphs above.
These statement-level issues can lead to unexpected failures during execution and are the primary cause of most of the failed cases we encountered.

The \emph{Functions} category refers to predefined functions not supported on the host. No SLT test case contains an unsupported function, while \DuckDBCompatibilityIssueFunction{} and \PostgreSQLCompatibilityIssueFunction{} issues in DuckDB and PostgreSQL test suites, respectively, are related to unsupported functions, such as functions for querying system information, text processing, or nested data. For example, in DuckDB, \texttt{SELECT range(3);} returns a three elements list \texttt{[0,1,2]}, while the other DBMSs do not support \texttt{range()}.

The \emph{Types} category refers to the donor-specific data types that are not supported in the host (\emph{e.g.}, nested data and big integer). 
\SQLiteCompatibilityIssueType{}, \DuckDBCompatibilityIssueType{}, and \PostgreSQLCompatibilityIssueType{} issues are related to types in the SQLite, DuckDB, and PostgreSQL test suites, respectively.
One issue that caused MySQL to fail to execute the DuckDB and PostgreSQL test cases is its requirement to specify the maximum length for the \texttt{VARCHAR} type.
Note that, SQLite had fewer issues in the Types category, as it has a dynamic type system,\footnote{\url{https://sqlite.org/flextypegood.html}} which, for example, allows users to store values of any data type in a column, even if the declared column does not match the value's type. This is the reason why SQLite achieves a higher success rate on the PostgreSQL and DuckDB test suites than others.

The \emph{Operators} category refers to unsupported donor-specific operators that either lead to unexpected syntax errors or encounter incompatible operand pairs.
Examples include the type cast operator (\texttt{::}) mentioned above, and the unsupported \texttt{+} operator between string and integer in PostgreSQL, which is supported in SQLite. 
We identified \SQLiteCompatibilityIssueOperator{}, \DuckDBCompatibilityIssueOperator{}, and \PostgreSQLCompatibilityIssueOperator{} issues in SQLite, DuckDB, and PostgreSQL test suite, respectively.

The \emph{Configurations} category refers to unsupported configuration variables in the corresponding host. For example, one DuckDB test file initially uses \texttt{SET default\_null\_order=`nulls\_first';} to change the setting for the order of null values. This statement failed in PostgreSQL because it is an unknown system setting. As a result, when executing subsequent test cases in the same test file, PostgreSQL is unable to return values in the expected order.

The \emph{Semantic} category refers to functions, operators, or statements that share the same name but exhibit semantic inconsistencies. Specifically, they return different values when executed under different DBMSs, leading to discrepancies in query results.
We observed \SQLiteCompatibilityIssueSemantic{}, \DuckDBCompatibilityIssueSemantic{} and \PostgreSQLCompatibilityIssueSemantic{} related issues in SQLite, DuckDB, and PostgreSQL test suite. Although SLT has a large number of failing tests due to semantic reasons, most of the cases have the same root cause. For example, all 104K failing cases for DuckDB are due to the inconsistent behavior for the \texttt{/} operator, which is a decimal division in DuckDB, but an integer division in SQLite.
Some DBMSs would be built compatibly with other DBMSs (\emph{e.g.}, DuckDB aims to partly match the semantics of PostgreSQL).
\squality{} could identify cases that exhibit inconsistencies, offering developers guidance, as these inconsistencies would not cause exceptions, but silently produce unexpected results, which are difficult to notice.
Listing~\ref{listing:example-duckdb-null} and Listing~\ref{listing:example-duckdb-pg} show differences in anonymous records sorting and eponymous functions (\emph{e.g.} PostgreSQL functions starting with ``pg\_'') respectively between DuckDB and PostgreSQL.
We reported the issue in Listing~\ref{listing:example-duckdb-null} to the developers, and after a discussion, they concluded to deliberately deviate from other DBMSs like PostgreSQL.
\begin{table}[tb]
    \centering
    \caption{Summary of test cases from each test suite that bring difficulties for reuse}
    \begin{tabular}{lrrr}
    \toprule
         & SQLite & DuckDB & PostgreSQL\\
    \midrule
    Dialect-specific features & 0.1\%  & 70.2\% & 72.7\%\\
    Syntax differences        & 12.8\% & 23.9\% & 26.4\%  \\
    Semantic differences      & 87.1\%  & 5.9\% & 0.9\% \\
    \bottomrule
    \end{tabular}
    \label{tab:my_label2}
\end{table}
\textcolor{black}{
\paragraph{Summary statistics}
We present summary statistics of test cases of each test suite in Table~\ref{tab:my_label2}. 
We manually classified the test cases based on the failures due to feature, syntax, and semantic differences. 
SLT is the most standard-compliant test suite, and thus only 12.9\% of the failures were due to unique features or deviations in syntax. 
Conversely, the PostgreSQL and DuckDB test suites consist of test cases for unique features, which caused most failures when executing test cases across different DBMSs.}
\textcolor{black}{\paragraph{Implications}
We summarize the below implications about the test suite compatibility.
First, it is difficult to reuse dialect-specific features, such as unique data types. 
Second, we believe that the syntax differences could be partially addressed by using SQL translators---potentially by using large language models---by converting SQL statements from one dialect to another to prevent errors.
Third, manual effort is necessary to account for semantic differences—such as functions, operators, or statements that share the same name but exhibit semantic inconsistencies. These differences may arise from developers’ different design choices or potential bugs in the system. Developer inspection of these cases is beneficial for identifying bugs and compatibility issues, as well as improving documentation. This effort is feasible, because semantic issues comprise only a small portion of all failures.
}
    

\begin{lstlisting}[language=SQL, float=tb, caption={The result value of DuckDB is \texttt{true}, whereas other DBMSs return \texttt{null}.}, label={listing:example-duckdb-null}, belowskip=0pt, aboveskip=0pt]
SELECT (null, 0) > (0, 0);
-- DuckDB: true
-- PostgreSQL: null
\end{lstlisting}
\begin{lstlisting}[language=SQL, float=tb,  caption={DuckDB always return \texttt{true} even if passing invalid arguments to this function.}, label={listing:example-duckdb-pg}, belowskip=0pt]
select has_column_privilege(1,1,1);
-- DuckDB: true
-- PostgreSQL: ERROR
\end{lstlisting}
\result{
We reused three test suites across four DBMSs, which we used to find 3 crashes and 3 hangs. Sampling 100 failing tests, we identified various compatibility issues: 18 in SQLite, 59 in DuckDB, and 53 in PostgreSQL, which can be attributed to SQLite's test cases exercising mostly standard functionality, while the latter two comprised more DBMS-specific test cases. 
}

\section{Related Work}

\paragraph{DBMS Testing}
Methods have been proposed for automatically identifying bugs in DBMSs in recent years. SQLancer creates databases and queries, and then autonomously validates the results provided by the DBMS~\cite{rigger2020detecting, rigger2020finding, rigger2020testing, ba2023testing}. 
Transformed Query Synthesis (TQS)~\cite{tang2023detecting} detected bugs of join optimization in DBMSs and TxCheck~\cite{jiang2023detecting} detected transactional bugs by constructing graph-based oracles.
APOLLO~\cite{jung2019apollo} identifies performance bugs by executing queries on both an older and a newer version of the same DBMS. 
Our test suite has the potential to augment the efficiency of these methods. In particular, it can be employed to serve as a basis for additional mutations. 
For example, Sedar~\cite{fu2024sedar} successfully detected crashes by reusing test cases from other DBMSs. However, in contrast to our work, Sedar only re-used the SQL statements, and not their results specification.

\paragraph{Studies on SQL features}
Several empirical studies have been conducted that closely relate to our study.
Cui et al.~\cite{cui2023differentially} studied transaction implementations and found transactional compatibility issues across DBMSs using a differential-testing methodology, while our study has a more general scope in identifying compatibility issues.
Gupta et al.~\cite{gupta2021procedural} investigated the use of procedural extensions in SQL---specifically, stored procedures, user-defined functions (UDFs), and triggers---in the context of Microsoft Azure SQL Database Service. 
They subsequently developed a benchmark suite and evaluated these procedural extensions, that is, the time/resources diverse procedural SQL components spent in a given workload. 
Toussaint et al.~\cite{toussaint2022troubles} conducted a survey on the usage of \texttt{NULL} values among DBMS users. They examined the drawbacks associated with \texttt{NULL} values in SQL and how they are handled.
Vogelsgesang et al.~\cite{vogelsgesang2018get} analyzed the characteristics of queries generated by a modern \emph{business intelligence} (BI) tool, and found that they differ from benchmarks used in the industry.
Both the above work and our study investigated how SQL is used in practice.
However, our study primarily focuses on the testing perspective. We have examined the SQL test suites, aiming to reuse the test cases in order to discover new bugs within the DBMSs and enhance their reliability.

\paragraph{Programming language standards}
Our study gave quantitative and qualitative evidence of how SQL dialects differ in practice. 
Various studies have been conducted to explore the standards of SQL and different programming languages, or formalize them.
Guagliardo et al.~\cite{guagliardo2017formal} addressed the lack of formal semantics for real-world SQL queries and provided formal semantics for a basic class of SQL queries. 
Memarian et al.~\cite{memarian2016into} provided an in-depth investigation into the semantics of pointers and memory in C. 
Furthermore, they extended their study concerning pointer provenance~\cite{memarian2019exploring}. 
They examined the discrepancies between the actual C standard and what expert C programmers believe the standard ensures, while we focus on discrepancies between different DBMSs.

\paragraph{Test reusability}
Various studies have been conducted on software reuse.
Tiwari et al.~\cite{tiwari2013reuse} argued the importance of software reuse and discussed approaches to test reuse. They mentioned test reuse could reduce the test effort.
Recently, test reuse has been proposed for UI testing. Zhao et al.~\cite{zhao2020fruiter} introduced a framework that could automatically evaluate UI test reuse. Mariani et al.~\cite{mariani2021semantic} conducted the first empirical study on GUI events' semantic matching and reported significant findings to enhance test reuse approaches.
Regression test suites have long been a resource for mutation-based testing. Le et al.~\cite{le2014compiler} found bugs in LLVM by using test cases from GCC. Zhong et al.~\cite{zhong2020squirrel} used the test suite of each DBMS as the seed corpus for their mutation-based fuzzing approach, which found many bugs in DBMSs.

\section{Threats to validity}
We used a common methodology~\cite{falessi2018empirical} to identify and mitigate potential threats to our work.
\paragraph{Internal validity}
The main threat to the internal validity of this work is potential inaccuracies in the manual classification of failed test cases in RQ3 and RQ4, which could introduce bias, leading to misclassification of the reasons for failed cases.
We alleviate this threat by carefully investigating the documentation and discussing any unclear reasons among the authors. Where we still failed to determine the reasons after discussion and studying the user manuals and implementation of the DBMSs, we reported the issue to DBMS developers. 
Thus, we expect that any misclassifications are unlikely and would have little impact on the overall results.

\paragraph{Construct validity}
The main threat to construct validity is that we designed our own test suite, whose design decisions might affect the analyses. For instance, we chose to execute all test cases using Python clients, and in Section~\ref{sec:Result-RQ3}, we found that test cases can compute different results depending on which client is used.
Implementing test case translators and using the original test runner (\emph{e.g.}, parsing DuckDB test cases to run on SQLite via the SLT runner) would not scale for a larger number of test suites, as every test runner would need to support every test case format.
Despite this potential threat, we believe that the high-level insights of the study are valid even when considering such different potential experimental setups.

\paragraph{External validity}
The external validity of our study might be constrained by the limited scope of the open-source DBMSs that we examined. 
We investigated four widely used open-source DBMS test suites and reused three of them. However, commercially developed systems, for example, Oracle Database, might have more extensive test suites since companies might have more resources.
To mitigate this issue, we have also investigated the test suite of CockroachDB---a DBMS mainly developed commercially (by Cockroach Labs)---and observed similar trends there.
Additionally, the evolution of DBMSs might lead to changes in the test suites we studied, potentially limiting the applicability of our results over time.
\section{Implications}

Based on our study, we have identified important actionable insights, which we subsequently summarize.

\paragraph{Overcoming scalability challenges}
While being the first systematic effort in test reuse of DBMSs, \squality{} has not fully addressed the issues of unifying SQL test suites. 
DBMS test suites contain a variety of different test runner commands that need to be supported in our design.
We believe that supporting various test case formats and runner commands, as attempted in our work, is non-scalable due to the amount of manual implementation effort that is involved.
One pragmatic, alternative approach to integrating various existing test suites based on different runner commands could involve removing all test runner commands from the test cases, executing them on the \emph{donor} system, and assuming the outputs as the ground truth. However, this approach would necessitate manual inspection of the results to validate whether they are correct. 

\paragraph{Test suites for new DBMSs}
For newly developed DBMSs, we recommend adopting the SQLite test format and test suite.
The reason for this is three-fold: (1) the test format is relatively simple (see Section~\ref{sec:Result-RQ1}), facilitating the implementation of the test runner required to execute the test suite on the DBMS; (2) SQLite's tests are mostly standard-compliant, with only few testing SQLite-specific functionality, allowing their reuse (see Section~\ref{sec:Result-RQ2}); and (3) compared to other systems, they have relatively few dependencies (see Section~\ref{sec:Result-RQ3}). \textcolor{black}{We found that, besides SQLite and DuckDB, also other DBMSs, such as MonetDB\footnote{\url{ https://github.com/MonetDB/MonetDB/tree/master/sql/test}} and Databend\footnote{\url{https://github.com/datafuselabs/databend}} use the SLT format in their test suites.}
As an alternative, for DBMSs designed to be (partly) compatible with a potential \emph{donor} DBMS, we advise utilizing its test suites, as they can find both bugs and compatibility issues (see Section~\ref{sec:eva-rq4}). 

\begin{table}[tb]
    \centering
    \caption{Coverage of executing each original test suite and SQuaLity on SQLite, DuckDB, and PostgreSQL}
    \vspace{-3mm}
    \small
    \begin{tabular}{lrrrrrr}
    \toprule
         & \multicolumn{2}{c}{SQLite} & \multicolumn{2}{c}{DuckDB} & \multicolumn{2}{c}{PostgreSQL}\\
         & Line & Branch & Line & Branch & Line & Branch \\
    \midrule
    Original tests & 26.9\% & 19.8\% & 72.8\% & 46.4\% & 62.1\% & 47.2\% \\
    SQuaLity       & 43.4\% & 34.5\% & 74.0\% & 47.2\% & 63.0\% & 48.2\% \\
    \bottomrule
    \vspace{-5mm}
    \end{tabular}
    \label{tab:coverage}
\end{table}

\textcolor{black}{\paragraph{Benefits of reusing test suites} Our study provides the below key insights for DBMS developers. 
First, SLT, which contains mainly standard-compliant SQL statements, can be adapted to different DBMSs with a higher success rate, and could thus help detect logic errors for basic features during development.\footnote{\url{https://github.com/dolthub/dolt/issues/7079}}
Second, other test suites often contain complex and DBMS-specific features that may restrict or even prohibit test case reuse; despite this, we observed that some features are shared across different systems and the test cases are valuable for finding potential bugs.
Third, reusing the composed test suite \squality{} can help increase test coverage compared to using only the original test suite, which we determined in an additional coverage experiment (see Table~\ref{tab:coverage}).
We explain this increase by the larger range of SQL statements and SQL features from different test suites that can improve coverage for features not covered in the existing test suite as well as by testing error handling of both parsing and semantic analysis when processing unsupported or partly-supported features.
SLT achieves a branch coverage of only 20\% on SQLite. This is due to SLT containing mostly standard-compliant SQL commands.
SQLite has three harnesses. 
For example, the 100\% branch coverage for SQLite is achieved by another test suite of SQLite, TH3,\footnote{\url{https://sqlite.org/th3.html}} which not only contains SQL test cases, but also test cases written in C. Although SQLite and SLT are open-sourced, the TH3 test suite is proprietary.
}

\textcolor{black}{\paragraph{Supporting a new DBMS}
Applying SQuaLity on a new DBMS requires supplementing the general test runner implementation with DBMS-specific logic, as also explained in our artifact.
Specifically, interfaces need to be implemented, which establish connections to the DBMS, set up an initial database and subsequently remove it, and execute statements and queries. 
Implementing them typically requires a low effort, as they are implemented in 33 LOC on average for the DBMSs in our experiments.
Executing SQuaLity on a new DBMS is likely to cause test failures, caused by different root causes.
First, crashes are never expected. Hangs, which are also abnormal, require developers to inspect them---they can be missed optimization or bugs.
Second, when the DBMS under test is designed for compatibility with existing dialects (\emph{e.g.}, PostgreSQL), developers are advised to inspect any inconsistencies provoked by the test cases from the DBMSs they aim to be compatible with.
The discrepancies could indicate a potential bug since the semantics should match. 
For failures in test cases with dialect incompatibilities, the possibility of finding bugs is lower, and such cases could be removed.
Third, developers can identify patterns in unexpected error messages to identify potential bugs. For example, messages starting with \texttt{``INTERNAL Error''} are never expected in DuckDB, and suggest bugs. Matching the patterns (\emph{e.g.}, using regular expressions) of the error messages in the execution log can help detect issues after running SQuaLity.}

\paragraph{Filling testing gaps}
Our study provides actionable insights on what features are undertested, warranting additional test cases to be written.
For example, the SQL statements distribution in RQ2 suggests that only few test cases exercise \texttt{WITH} statements (\AvgWithProportion{}), which are used to specify CTEs.
In particular, we found a bug in MySQL (see Listing~\ref{listing:mysql-crash}), which was identified and assigned CVE-2024-20962, due to an issue caused by a recursive CTE. 
This suggests that this feature, and other statements for which only few test cases exist, might be insufficiently tested.

\paragraph{Understanding SQL dialects}
While SQL is commonly seen as a single language, various SQL dialects exist as our experiments indicate. 
For example, previous research~\cite{guagliardo2017formal} on formalizing SQL semantics has focused on general semantics, rather than accounting for different SQL dialects. 
We identified various incompatibility issues when executing test suites across different DBMSs---18 in SQLite, 59 in DuckDB, and 53 in PostgreSQL (see RQ4)---due to different reasons (\emph{e.g.}, incompatible statements, operators, and types).
These results might inform variability points for formal semantics.

\paragraph{Enhancing automated testing}
Many DBMS fuzzing approaches require a seed corpus, based on which they derive follow-up inputs through mutation, and our test suite---containing more than 7 million SQL test cases---could be used as such a corpus. 
Recent work has demonstrated the benefits of this.
For example, Sedar~\cite{fu2024sedar} has detected bugs in well-known DBMSs by reusing SQL test cases. 
\section{Conclusion}
In this paper, we have contributed a unified test suite consisting of more than 7 million SQL statements from three test suites of widely used DBMSs as well as an empirical study concerning different test suite features, test case patterns, and testing dependencies. 
Our findings for test suite reuse are encouraging.
We identified crashes and hangs by executing test cases written for one DBMS on other DBMSs, indicating the benefits of test-case reuse.
However, we also identified various challenges that complicate test reuse.
This includes the feature variety in test suite formats, test runner commands (up to 112 commands, RQ1), SQL dialects (up to 31\% of the test cases in PostgreSQL are not standard-compliant, RQ2), and environmental dependencies (such as file path and locale configurations, RQ3).
We hope that this first study toward test case reuse for DBMSs will inspire follow-up work and inform practitioners of the opportunities and challenges of reusing test cases.



\section*{Acknowledgements}
This research is supported by the Ministry of Education, Singapore, under the Academic Research Fund Tier 1 (FY2023) as well as the National Research Foundation, Singapore, and Cyber Security Agency of Singapore under its National Cybersecurity R\&D Programme (Fuzz Testing). Any opinions, findings and conclusions, or recommendations expressed in this material are those of the author(s) and do not reflect the views of National Research Foundation, Singapore, and Cyber Security Agency of Singapore.

\balance
\bibliographystyle{ACM-Reference-Format}
\bibliography{sample-base}

\received{April 2024}
\received[revised]{July 2024}
\received[accepted]{August 2024}
\end{document}